%% file: acl_latex.tex
\pdfoutput=1

\documentclass[11pt]{article}

\usepackage[preprint]{acl}

\usepackage{times}
\usepackage{latexsym}

\usepackage{multirow}
\usepackage{adjustbox}
\usepackage{booktabs} 
\usepackage{amsmath}  
\usepackage{graphicx}
\usepackage{subcaption}
\usepackage{caption}
\usepackage{amssymb}
\usepackage[shortlabels]{enumitem}
\usepackage{tcolorbox}
\usepackage{array}
\usepackage{pifont}
\usepackage{colortbl}
\usepackage{xcolor}
\usepackage{wasysym}
\usepackage{hyperref}

\usepackage[T1]{fontenc}

\usepackage[utf8]{inputenc}

\usepackage{microtype}

\usepackage{inconsolata}

\usepackage{graphicx}
\usepackage{booktabs}
\usepackage{array}
\usepackage{xcolor}
\usepackage{soul}
\usepackage{natbib}
\newcommand{\hlc}[2][yellow]{{\sethlcolor{#1}\hl{#2}}}
\definecolor{pink}{RGB}{255,192,203}
\definecolor{cyan}{RGB}{0,255,255}
\definecolor{green}{RGB}{0,255,0}
\definecolor{darkred}{rgb}{0.7215686274509804, 0.2235294117647059, 0.27058823529411763}
\definecolor{lightblue}{RGB}{129, 209, 241}
\definecolor{new_red}{RGB}{248, 157, 134}

\definecolor{mygreen}{RGB}{0,160,0}

\definecolor{myred}{RGB}{178,34,34}
\definecolor{my_lightred}{rgb}{1.0, .941, .902}

\newcommand{\cmarkgreenb}{\textcolor{mygreen}{{ \ding{51}}}}

\usepackage{amsfonts}
\usepackage{bm}
\title{UniHGKR: Unified Instruction-aware Heterogeneous Knowledge Retrievers}

\author{
 \textbf{Dehai Min\textsuperscript{1}},
 \textbf{Zhiyang Xu\textsuperscript{3}},
 \textbf{Guilin Qi\textsuperscript{1}},
 \textbf{Lifu Huang\textsuperscript{4}},
 \textbf{Chenyu You\textsuperscript{2}}
\\
\\
 \textsuperscript{1}Southeast University,
 \textsuperscript{2}Stony Brook University,
 \textsuperscript{3}Virginia Tech,
 \textsuperscript{4}UC Davis
\\
\\
\texttt{qieqiemin@gmail.com, zhiyangx@vt.edu, gqi@seu.edu.cn}
\\
\texttt{lfuhuang@ucdavis.edu, chenyu.you@stonybrook.edu}
}

\begin{document}
\maketitle
\input{sections/abstract}

\input{sections/introduction}

\input{sections/relatedwork}

\input{sections/dataset}

\input{sections/UniHGKR}

\input{sections/experimental_method}

\input{sections/result}

\input{sections/conclusion}

\input{sections/limitations}


\bibliography{custom}


\appendix

\input{sections/appendix}

\end{document}

%% file: sections/abstract.tex
\begin{abstract}
Existing information retrieval (IR) models often assume a homogeneous structure for knowledge sources and user queries, limiting their applicability in real-world settings where retrieval is inherently heterogeneous and diverse.
In this paper, we introduce UniHGKR, a unified instruction-aware heterogeneous knowledge retriever that (1) builds a unified retrieval space for heterogeneous knowledge and (2) follows diverse user instructions to retrieve knowledge of specified types. 
UniHGKR consists of three principal stages: heterogeneous self-supervised pretraining, text-anchored embedding alignment, and instruction-aware retriever fine-tuning, enabling it to generalize across varied retrieval contexts. This framework is highly scalable, with a BERT-based version and a UniHGKR-7B version trained on large language models. 
Also, we introduce CompMix-IR, the first native heterogeneous knowledge retrieval benchmark. It includes two retrieval scenarios with various instructions, over 9,400 question-answer (QA) pairs, and a corpus of 10 million entries, covering four different types of data.
Extensive experiments show that UniHGKR consistently outperforms state-of-the-art methods on CompMix-IR, achieving up to 6.36\% and 54.23\% relative improvements in two scenarios, respectively.
Finally, by equipping our retriever for open-domain heterogeneous QA systems, we achieve a new state-of-the-art result on the popular ConvMix~\cite{christmann2022conversational} task, with an absolute improvement of up to 5.90 points.\footnote{Our code, datasets and model checkpoints are available at: \href{https://github.com/ZhishanQ/UniHGKR}{https://github.com/ZhishanQ/UniHGKR}}

\end{abstract}


%% file: sections/introduction.tex
\section{Introduction}

Retrieval-Augmented Generation (RAG~\cite{2020RAG,gao2023retrieval,qi2024roravlmrobustretrievalaugmentedvision}) has become a pivotal technique for improving the faithfulness of generative large language models (LLMs~\cite{achiam2023gpt}). By leveraging retrievers to extract relevant knowledge from large-scale knowledge corpus, RAG effectively reduces the hallucinations often produced by LLMs \cite{ayala2024reducing,muennighoff2024generative}.

\begin{figure}[t]
    \centering
    \begin{subfigure}[b]{0.48\textwidth}
        \centering
        \includegraphics[width=\textwidth]{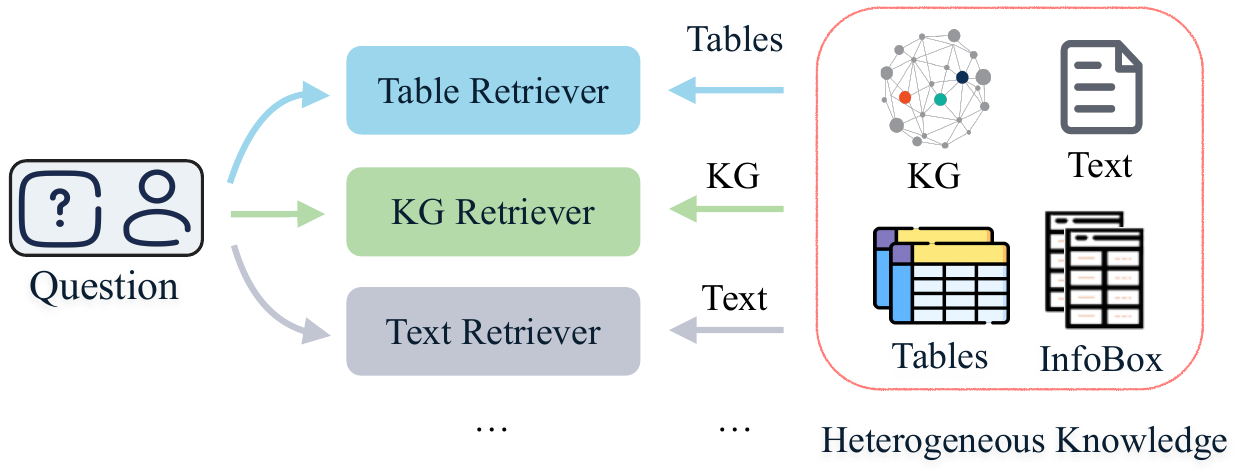}
        \caption{Conventional retrievers focus on a single data type.}
        \vspace{2mm}
    \end{subfigure}
    
    \begin{subfigure}[b]{0.48\textwidth}
        \centering
        \includegraphics[width=\textwidth]{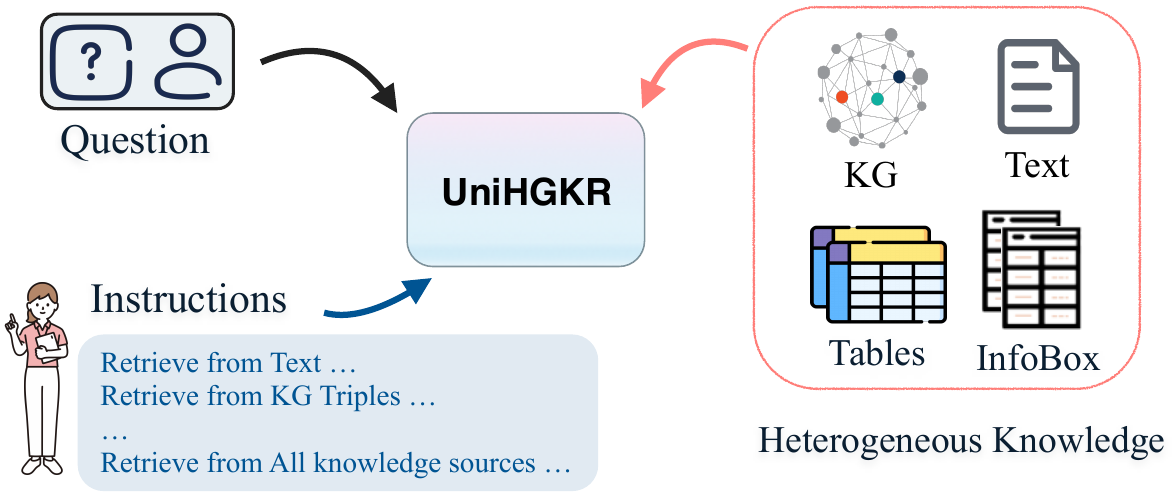}
        \caption{UniHGKR aims to retrieval from any heterogeneous knowledge source.}
    \end{subfigure}
    \caption{Compared to traditional methods, UniHGKR follows user instructions to process queries and retrieves from a heterogeneous knowledge candidates pool.}
    \label{fig:figure_1}
    \vspace{-4mm}
\end{figure}

Although existing information retrieval (IR) methods\cite{yang2024trisampler, zhao2024dense} have demonstrated effectiveness in retrieving information from homogeneous knowledge corpus, where knowledge is stored in a single structure, such as tables~\cite{kong2024opentab} or text~\cite{behnamghader2024llm2vec}, most of these systems fail to recognize diverse user retrieval intents and retrieve heterogeneous knowledge from multiple sources. In heterogeneous IR, knowledge comes from multiple structures, making retrieval much more complex. Relying solely on homogeneous knowledge often results in partial or incomplete retrieval results, limiting the applicability of these systems to a wider range of downstream tasks \cite{asai2023task, christmann2022conversational}. For example, a retriever specialized in table-based retrieval \cite{herzig2021open} cannot be easily applied to downstream tasks such as question answering (QA) based on knowledge graphs~\cite{huang2023question}.

In this paper, we propose the Unified HeteroGeneous Knowledge Retriever (\textbf{UniHGKR}), a novel framework designed to retrieve information from heterogeneous knowledge corpus by following user instructions, as depicted in Figure \ref{fig:figure_1}. The UniHGKR framework consists of three training stages: \textbf{(1) Unified Embedding Self-Supervised Pre-training}: This stage addresses the lack of structured data in the original pretraining of the language model, laying the foundation for the creation of a unified embedding space. \textbf{(2) Text-Anchored Heterogeneous Embedding Alignment}: In this stage, natural language text that shares the same semantic content as heterogeneous data is collected, and their embeddings are aligned using contrastive learning. This process creates a unified embedding space that captures semantic information, independent of the format in which the knowledge is presented. \textbf{(3) Instruction-Aware Heterogeneous Retriever Fine-tuning}: At this final stage, the retriever is fine-tuned on heterogeneous knowledge retrieval tasks. To enhance the model's capability to follow user instructions, we introduce two specialized contrastive losses, termed `type-balanced loss' and `type-preferred loss', which are designed to optimize retrieval performance according to user instructions.

In addition, existing heterogeneous IR benchmarks have limited knowledge coverage~\cite{petroni2021kilt,muennighoff2023mteb}. For example, studies like \cite{chen2021open,zhong2022reasoning} focus only on two types of knowledge: tables and text. To address this gap, we introduce ~\textbf{CompMix-IR}, the first-ever benchmark for heterogeneous knowledge retrieval. CompMix-IR has over 9,400 QA pairs and a corpus of 10 million entries spanning four distinct knowledge types: Text, Knowledge Graphs (KG), Tables, and Infoboxes. Derived from the open-domain QA dataset CompMix \cite{christmann2024compmix}, CompMix-IR transforms this QA task into a standard IR task (as detailed in Section~\ref{sec:CompMi_IR}). To better reflect real-world retrieval needs, we define two distinct scenarios in this benchmark: (1) retrieving relevant evidence across all knowledge types, and (2) retrieving evidence of a specific type, as specified by user instructions. Both scenarios utilize the same evidence pool, requiring the retriever to adapt query-evidence similarity based on the instructions. This setup mirrors the complexities of real-world retrieval tasks, offering enhanced practical relevance and utility for diverse applications.


Experimental results demonstrate the effectiveness of our proposed UniHGKR over the existing methods, with relative improvements of up to 6.36\% and 54.23\% in two different scenarios. In addition to the BERT-based UniHGKR-base model, we also extend our framework to an LLM-based retriever and train the UniHGKR-7B model to verify scalability. Both models achieve state-of-the-art (SOTA) performance on CompMix-IR respective to their parameter scales.
Furthermore, in the context of open-domain heterogeneous QA, systems equipped with UniHGKR retriever set a new SOTA on the ConvMix task~\cite{christmann2022conversational}, with an absolute gain of up to 5.90 points, further validating its real-world applicability.

%% file: sections/relatedwork.tex
\section{Related Work}

\noindent \textbf{IR on Heterogeneous Knowledge.}
Several efforts have been make in this field, but they come with notable limitations. For example, \citet{li2021dual,kostic2021multi} create separate retrieval indices for different data types, retrieving them individually. 
This approach fails to compare relevance of evidence across knowledge sources, and maintaining multiple indices increases system complexity. On the other hand, UDT-QA \cite{ma2022open_2} introduces a verbalizer-retriever-reader framework, using a finetuned data-to-text generator \cite{nan2021dart} to convert heterogeneous scenarios into homogeneous text scenarios. However, this leads to answer coverage loss and limits downstream reader models from utilizing the structured of data, essential for tasks like KG-based and Table-based QA \cite{hu2023empirical,kweon2023open}. Additionally, these retrievers are typically designed for predefined single tasks, failing to accommodate users diverse retrieval needs.

\noindent \textbf{QA over Heterogeneous Knowledge.}
Each data type has its own characteristics and provides unique benefits. Some studies explores the integration of knowledge sources to QA \cite{ma2022open, min-etal-2024-exploring,you2020towards,you2020contextualized,you2021self,you2021knowledge,you2021mrd,chen2021self,you2022end}. For instance, HybridQA~\cite{chen2020hybridqa} and OTT-QA~\cite{chen2021open} investigate the task of extracting answers from the combination of tables and text.
Going further, CONVINSE~\cite{christmann2022conversational}, Explaignn~\cite{christmann2023explainable} and FAITH~\cite{jia2024faithful} consider four knowledge sources like this paper. However, their primary focus is on the answer generation parts of the system. Their retrieval approach is a time-consuming online pipeline: identifying entity IDs in questions, then conducting online searches in Wikipedia and Wikidata~\cite{vrandevcic2014wikidata}, and finally employing BM25~\cite{robertson2009probabilistic} to rank a small set of evidence.

%% file: sections/dataset.tex
\section{CompMix-IR Benchmark}
\label{sec:CompMi_IR}
In this section, we provide a detailed description of the construction of CompMix-IR, the definition of retrieval scenarios, and their instruction schema.

\subsection{Heterogeneous Knowledge Collection}
We introduce \textbf{CompMix-IR}, the first native heterogeneous knowledge retrieval dataset, built on the CompMix dataset \cite{christmann2024compmix}, a recent crowdsourced open-domain QA task spanning four knowledge sources. However, the original dataset lacks a heterogeneous corpus suitable for retrieval tasks. To address this, we construct a heterogeneous knowledge corpus related to the CompMix QA set, extending it for IR tasks. Specifically, we collect and store four types of knowledge using the following methods for each question:

\begin{itemize}[topsep=1pt, itemsep=0.8pt, leftmargin=.1in, parsep=0pt]

\item \textbf{KG facts.} We use CLOCQ \cite{christmann2022beyond} to retrieve the top-1000 KG triples related to each question from the Wikidata dump. We also store the disambiguations and wikidata entities information returned by CLOCQ. This information helps us evaluate the relevance between the evidence and the question. To feed the structured data into the language model, the retrieved KG facts are linearized, with entities and relations separated by commas.
\item \textbf{Text, Tables and Infoboxes.} 
We use the entities mentioned in questions to retrieve the corresponding Wikipedia pages. Subsequently, a parser is used to extracts natural language paragraphs (text evidence), tables, and infoboxes from the pages. Also, we utilize hyperlinks from Wikipedia pages to map the corresponding entity mentions to Wikidata IDs. This achieves the same labeling format as KG evidence. Following \cite{oguz2022unik}, both tables and infoboxes are linearized using simple templates. Specifically, we concatenate the properties and values from the table using the word "is". The entity name described by the infobox and the properties and values are strung together by a comma ",", forming a text string. Additionally, Wikipedia page titles are added at the beginning of the evidence for clearer information.

\end{itemize}

\input{table/compmix_ir_statistics}

To align with the standard IR task setup, we use automated scripts to label relevant evidence (golden labels) for each question. The relevance between the evidence and the question is of a boolean type (True/False). Specifically, if the entities in the evidence contain the answer to the question, the relevance is marked as True; otherwise, it is marked as False. Each question has at least one piece of evidence that provides the answer.
The evidence retrieved for all questions in CompMix is combined into a heterogeneous knowledge pool, forming the corpus for the IR task. This corpus includes over 10 million pieces of evidence, covering knowledge about 137,808 different entities. Detailed statistics of CompMix-IR are presented in Table~\ref{tab:compmix_ir_statistics}, and examples of linearized evidence and QA pair, as well as their annotation information examples, provided in Appendix~\ref{sec:appendix_compmix_ir}.

\input{table/instructions}

\subsection{Retrieval Scenarios and Instructions} 

To address real-world heterogeneous knowledge retrieval needs, we define two distinct retrieval scenarios:
\begin{itemize}[topsep=1pt, itemsep=0.8pt, leftmargin=.1in, parsep=0pt]
\item \textbf{Scenario 1:} retrieving evidence from all types of knowledge. 
\item \textbf{Scenario 2:} retrieving type-specific evidence, as instructed by the user.
\end{itemize}

Both scenarios use the same evidence pool, requiring retrievers to consider not only the relevance of candidates but also whether these candidates match the data type specified in the instructions. Based on these two scenarios, we define an instruction schema (as shown in Table \ref{tab:instructions}), inspired by \cite{asai2023task,wei2023uniir}. Users can customize retrieval by adjusting the \textit{[domain]} and \textit{[source]} options, where \textit{[domain]} specifies the topic of evidence and \textit{[source]} defines the type of knowledge. Instructions are categorized into five groups: 
\(I_{\text{All}}\), \(I_{\text{Text}}\), \(I_{\text{KG}}\), \(I_{\text{Table}}\), and \(I_{\text{Info}}\). Here, \(I_{\text{All}}\) corresponds to our retrieval scenario 1, while the others correspond to scenario 2.
Additionally, to enhance the robustness of the instructions, each instruction was rewritten into 20 different expressions with the help of GPT-4o-mini~\cite{GPT_4o_mini}.

\begin{figure*}[ht]
\centering
     \includegraphics[width=0.98\textwidth]{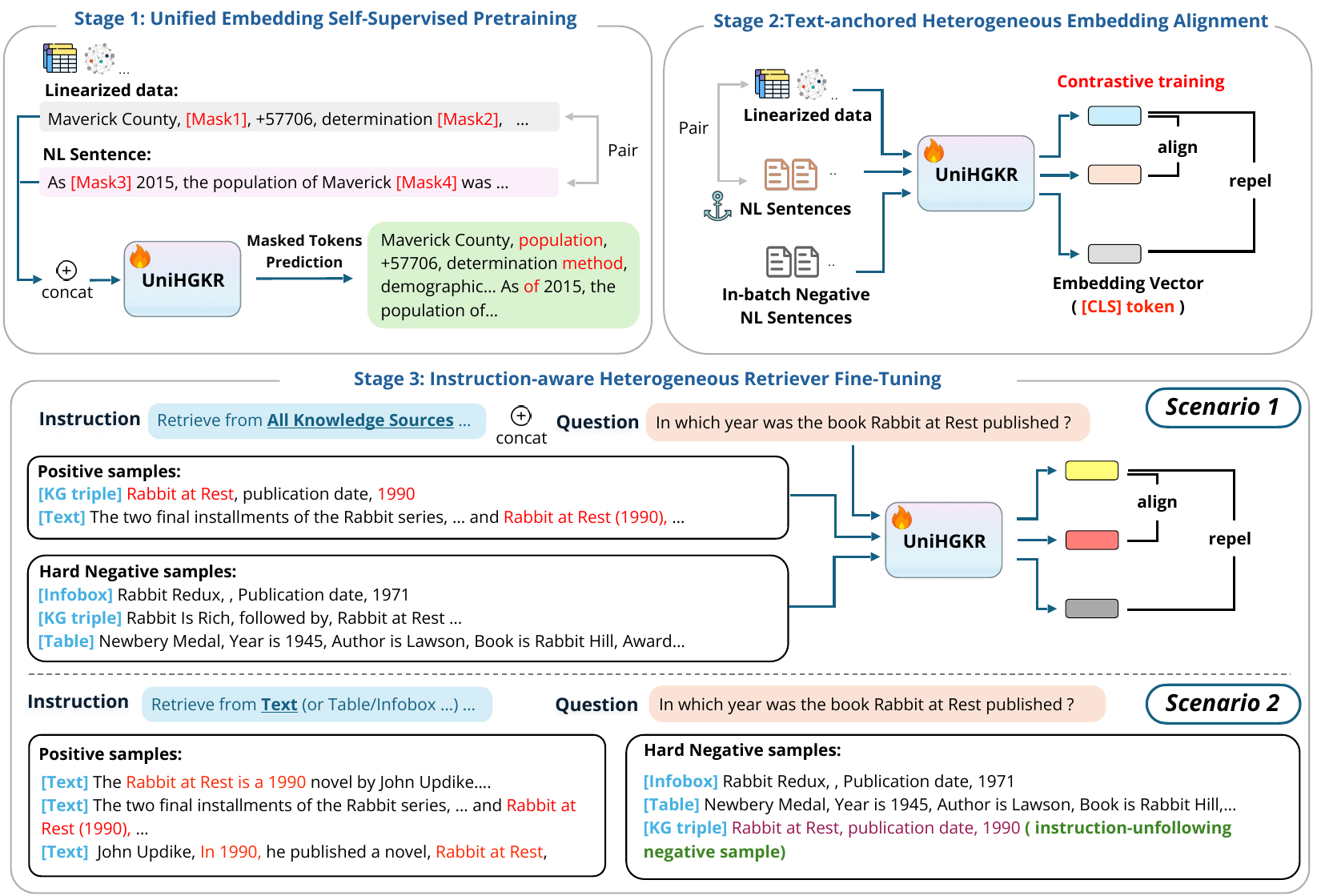}
     \caption{Illustration of our UniHGKR training framework.}
     \label{fig:figure_2}
     \vspace{-4mm}
\end{figure*}

%% file: table/compmix_ir_statistics.tex
\begin{table}[ht]
\centering\resizebox{0.44\textwidth}{!}
{

\begin{tabular}{lcrc} 
\hline
\textbf{Types} & \textbf{Avg. length} & \multicolumn{1}{c}{\textbf{Count}} & \textbf{Percentage} \\
\hline
Text & 19.86 & 5,916,596 & 57.74\% \\
KG & 11.40 & 2,214,854 & 21.61\% \\
Table & 20.32 & 1,043,105 & 10.18\% \\
Infobox & 11.05 & 1,072,440 & 10.47\% \\
Sum & 17.18 & 10,246,995 & 100.00\% \\
\hline
\end{tabular}

\vspace{-3mm} 

}
\caption{Statistics of CompMix-IR. `Avg. length' refers to the average number of words.}
\label{tab:compmix_ir_statistics}
\vspace{-4mm}
\end{table}

%% file: table/instructions.tex
\begin{table*}[th]
\renewcommand{\arraystretch}{1.2}
\setlength{\tabcolsep}{2pt}
\footnotesize
    \centering\resizebox{0.98\textwidth}{!}{
    \begin{tabular}{ll}
\toprule
\textbf{Template} & \text{Given a question in the \hlc[cyan!30]{[domain]} domain, retrieve relevant evidence to answer the question from the \hlc[green!30]{[source]}.} \\\midrule
\text{\hlc[cyan!30]{[domain]} options:\ \ } & \text{books, movies, music, television series, and football}  \\
\text{\hlc[green!30]{[source]} options:} & \text{All Knowledge Sources, Knowledge Graph Triples, Infobox, Table, and Text}  \\
\addlinespace
\hline
\text{Example 1:} & \text{Given a question in the \hlc[cyan!30]{music} domain, retrieve ... from \hlc[green!30]{Knowledge Graph Triples}.} \\
\text{Example 2:}& \text{Given a question in the \hlc[cyan!30]{football} domain, retrieve relevant ... from \hlc[green!30]{All Knowledge Sources}.} \\
\addlinespace
\hline
\text{Paraphrased 1:} & \text{For a question related to the \hlc[cyan!30]{music} domain, find pertinent information from \hlc[green!30]{Knowledge Graph Triples}.} \\
\text{Paraphrased 2:} & \text{For a question in the \hlc[cyan!30]{football} domain, extract helpful ... to address it from \hlc[green!30]{All Knowledge Sources}.} \\
\bottomrule
 \end{tabular}}
    \caption{Schema and examples of instructions for heterogeneous retrieval. The template contains two placeholders: \hlc[cyan!30]{[domain]} and \hlc[green!30]{[source]}. Users can select options for these based on their specific needs.}
    \label{tab:instructions}
\vspace{-2mm}
\end{table*}

%% file: sections/UniHGKR.tex
\section{UniHGKR}
\label{sec:UniHGKR}
In this section, we introduce our problem formulation and the UniHGKR framework. Our UniHGKR-base model adopts a single shared-encoder architecture, with parameters initialized from the BERT-base model~\cite{devlin-etal-2019-bert}. The [CLS] token from the final hidden layer is trained to serve as the embedding, following \cite{karpukhin2020dense,xiao2022progressively}.

\begin{figure*}[t]
\centering
     \includegraphics[width=0.98\textwidth]{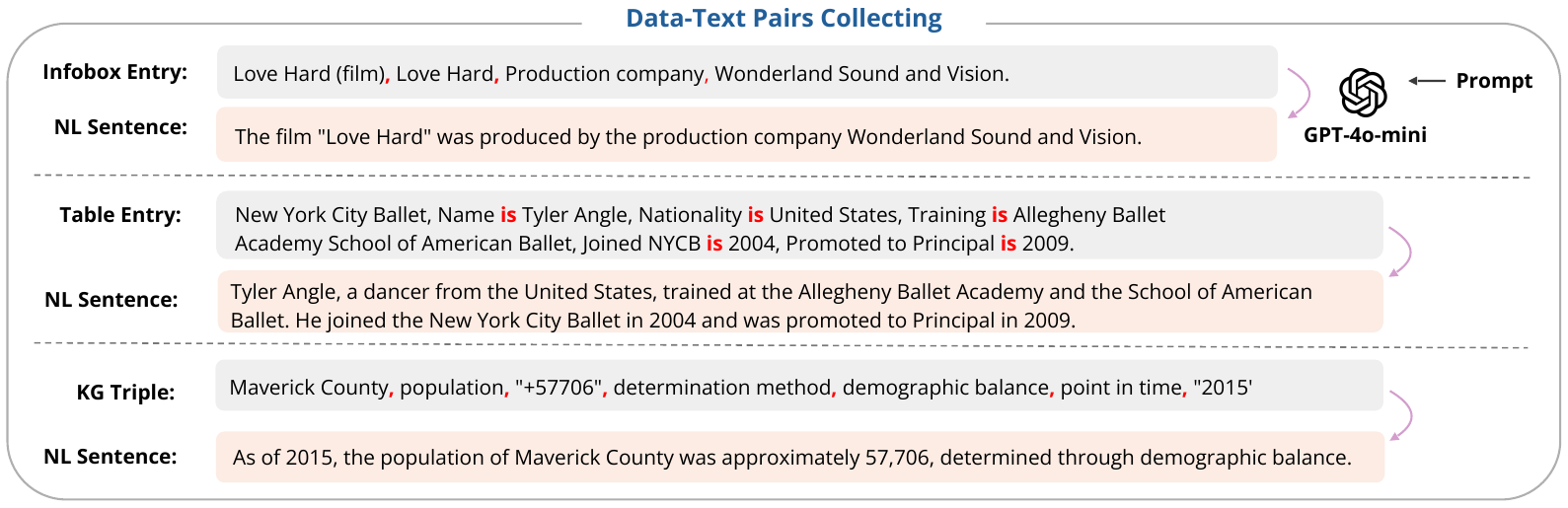}
     \caption{Illustration of Data-Text Pair Collection. The bold red \textbf{\textcolor{red}{is}} and the comma \textbf{\textcolor{red}{,}} are used in concatenation template when linearizing structured data. The prompts used for GPT-4o-mini can be found in Appendix \ref{sec:prompt_tempplate}.}
     \label{fig:figure_3}
     \vspace{-2mm}
\end{figure*}

\subsection{Problem Formulation}

Given a vase candidate pool of heterogeneous evidence \(\mathcal{E}\), defined as: 
\(\mathcal{E} = \bigcup_{\tau \in \mathcal{H}} \mathcal{E}_{\tau},\)
where \(\mathcal{H} = \{\text{Text},\ \text{Info},\ \text{Table},\ \text{KG}\}\) represents the set of evidence types. For each type \(\tau\), \(\mathcal{E}_{\tau} = \{e_{\tau}^i\}_{i=1}^{N_{\tau}}\) is the set of evidence of type \(\tau\).
The problem of retrieval with instructions is to find evidence \(e \in \mathcal{E}\) that is relevant to \(q\) according to the instruction \(I\).
The instruction and question are concatenated as \(\tilde{q} = [I; q]\), and the evidence \(e\) is encoded into embedding vectors by a shared encoder, denoted as \(\text{Enc}\).
The similarity between \(\tilde{q}\) and \(e\), is calculated as follows:
\[
f(\tilde{q}, e) = \text{Enc}\left(\tilde{q}\right)^\top \text{Enc}(e), \tag{1}
\]
where \({}^\top\) denotes the transpose operation. The retriever returns the top \(k\) evidence with the highest similarity as the retrieval results.

\subsection{UniHGKR Framework}
\label{sec:UniHGKR_train}
An overview of our framework is presented in Figure \ref{fig:figure_2}, which comprises the following three training stages:

\vspace{2mm}

\noindent \textbf{Stage 1: Unified Embedding Self-Supervised Pretraining.}
Pretrained Language Models (PLMs) are primarily trained on text, making them ineffective at generating embeddings for heterogeneous data, which is critical for IR tasks~\cite{li2022coderetriever,li2023structure}.
To this end, we design this stage to train PLMs with a token masking reconstruction task on heterogeneous data-text pairs as inputs. Specifically, we first construct a set of data-text pairs based on the CompMix-IR corpus with the help of LLMs, as illustrated in Figure~\ref{fig:figure_3}:
\[
\mathcal{D} = \left\{ \langle d_i, t_i \rangle \mid d_i \in \hat{\mathcal{E}},\ t_i = \mathcal{F}(d_i) \right\}_{i=1}^{N},\tag{2}
\label{eq:data_text_pairs}
\]
where, \(\hat{\mathcal{E}} = \mathcal{E}_{\text{KG}} \cup \mathcal{E}_{\text{Table}} \cup \mathcal{E}_{\text{Info}} \), \(\mathcal{F}\) is the data-to-text generator, which in our setting is GPT-4o-mini. 
The \(d_i\) is the linearized structured data, and the text \(t_i\) is a well-written natural language sentences with the same semantic information as \(d_i\). At this stage, they are concatenated to form training inputs. This approach enables the model to accept input sequences in heterogeneous formats as self-supervised signals. Furthermore, \(d_i\) and \(t_i\) can serve as distant supervision signals for each other, providing an indirect supervisory signal that enhances the model's learning from heterogeneous inputs~\cite{sun2021ernie,mintz2009distant}.
We adopt the token masking reconstruction task from RetroMAE~\cite{xiao2022retromae}: an additional single-layer Transformer~\cite{NIPS2017_3f5ee243} as a temporary decoder with a 50\% masking ratio, while our model serving as the encoder with a 15\% masking ratio. The training objective is:
\[
\min_{\theta} \sum_{x \in \mathcal{X}} -\log \mathrm{Dec}\left(x \mid \mathrm{Enc}\left(\tilde{x}; \theta\right); \theta\right). \tag{3}
\]
Here, \(x\) represents the original clean input, and \(\tilde{x}\) denotes the masked input.
After this stage training is completed, only the weights of the encoder are retained for subsequent training.

\vspace{2mm}
\noindent \textbf{Stage 2: Text-anchored Heterogeneous Embedding Alignment.}
Given that user instructions and questions are typically in text form, we further leverage the collected data-text pairs to optimize the embedding space anchored in text embedding representations. We apply contrastive learning~\cite{chen2020simple} to align the embedding of structured data \(d_i\) and text \(t_i\) that convey the same semantic information but differ in expression. Meanwhile, we repel embedding with different semantic information using in-batch negative samples \(B^-\) (samples that do not share semantic similarity with \(d_i\)) \cite{sohn2016improved}. This results in a unified embedding space focused on semantic information rather than the form of knowledge representation. The training objective is to minimize:
\[
\sum_{\langle d_i, t_i \rangle \in \mathcal{D} } -\log \frac{e^{f(d_i, t_i) / \tau}}{e^{f(d_i, t_i) / \tau} + \sum\limits_{b^- \in B^-} e^{ f(d_i, b^-) / \tau}},
\tag{4}
\]
where \(f()\) is a similarity function and \(\tau\) is the temperature parameter.

\noindent \textbf{Stage 3: Instruction-aware Heterogeneous Retriever Fine-Tuning.}
In this stage, we fine-tune our retriever on the heterogeneous knowledge retrieval task.
For each question \(q\) and its golden evidence \(e^+\), we generate two training samples: \((I_{\text{All}}, q, e^+)\) and \((I_{\lambda}, q, e^+)\), where \(\lambda\) is the data type of the positive sample \(e^+\). Additionally, we use the BGE model~\cite{xiao2024c} to mine hard negative samples set, denoted as \(E^-\).
For the contrastive training loss \(\mathcal{L}\):
\begin{equation}
\begin{aligned}
\mathcal{L} &= -\log \frac{e^{f(\tilde{q}, e^+) / \tau}}{e^{f(\tilde{q}, e^+) / \tau} + \sum_{d^- \in E^-} e^{f(\tilde{q}, e^-) / \tau}} \\
&= -\underbrace{f(\tilde{q}, e^+)/ \tau}_{\mathcal{L}_{\text{align}}} + \underbrace{\log \left( e^{f(\tilde{q}, e^+) / \tau} + \mathcal{L}_{\text{repel}}\right)}_{\mathcal{L}_{\text{uniformity}}} \\ 
\end{aligned}
\tag{5}
\end{equation}

\noindent Here, \(\mathcal{L}_{\text{align}}\) is the alignment loss encouraging higher similarity between the query and the positive evidence. Meanwhile, \(\mathcal{L}_{\text{uniformity}}\) denotes the uniformity loss applied over all samples, aiming to push the query away from negative samples~\cite{wang2020understanding}. We can simplify \(\mathcal{L}_{\text{repel}}\) :
\[
\mathcal{L}_{\text{repel}} = \sum_{\tilde{\lambda} \in \mathcal{H}} \ \sum_{e_{\tilde{\lambda}}^- \in E_{\tilde{\lambda}}^-} e^{f(\tilde{q}, e_{\tilde{\lambda}}^{-}) / \tau}
\tag{6}
\]

\noindent where \(\mathcal{H} = \{\text{Text},\ \text{Info},\ \text{Table},\ \text{KG}\}\), and \(E_{\tilde{\lambda}}^-\) is the set of hard negative samples of type \(\tilde{\lambda}\). We define: \(k_{\tilde{\lambda}} = |E^-_{\tilde{\lambda}}|, \tilde{\lambda} \in \mathcal{H}\) to represent the number of negative samples for each type.

To enhance the model's ability to follow user instructions, we design distinct contrastive losses: a type-balanced loss \(\mathcal{L}_{\text{balanced}}\) for training samples wtih instruction \(I_{\text{All}}\) (Scenario 1), and a type-preferred loss \(\mathcal{L}_{\text{preferred}}\) for training samples with instruction \(I_{\lambda}\) (Scenario 2). Specifically, for type-balanced loss \(\mathcal{L}_{\text{balanced}}\), we make \(k_{\text{Text}} \approx k_{\text{Info}} \approx k_{\text{Table}}  \approx k_{\text{KG}}\) depend on their numbers in \(E^-\). 
In contrast, for type-preferred loss \(\mathcal{L}_{\text{preferred}}\), in order to make the model learn the priority of evidence with specified-type \(\lambda\), we deliberately make \(k_{\lambda}\) significantly lower than the quantity of other types.
For example, when a training sample with instructions \(I_{\text{Table}}\), we set \(k_{\text{Text}} \approx k_{\text{Info}} \approx k_{\text{KG}} > k_{\text{Table}} = 0\), by filtering out \(e_{\text{Table}}^-\) from \(E^-\). 
By adjusting \(k_{\lambda}\), the training samples with \(I_{\lambda}\) have fewer negative samples of type \(\lambda\), thereby forming a preference for evidence of type \(\lambda\) in the global heterogeneous candidate pool.
Since we also use in-batch negative samples \(B^-\) during training,  the model can still learn to repel \(e_{\lambda}^-\), which are of the correct type but irrelevant evidences. Additionally, we also add a small number of instruction-unfollowing negative samples, which are related to \(q\) but not of the type \(\lambda\), to encourage the model to decrease their similarity with \(\tilde{q}\).



%% file: sections/experimental_method.tex
\section{Experimental Methodology}

In our main experiments, we train and evaluate retrievers on the CompMix-IR, following the train, dev, and test set divisions in CompMix.

\subsection{Baselines}

\noindent \textbf{Zero-shot SoTA Retriever.} Referring to the MTEB leaderboard\footnote{\href{https://huggingface.co/spaces/mteb/leaderboard}{https://huggingface.co/spaces/mteb/leaderboard}}, we select some top-ranking and SOTA models as baselines, including Mpnet~\cite{song2020mpnet}, Contriever~\cite{izacard2022unsupervised}, DPR~\cite{karpukhin2020dense}, GTR-T5~\cite{ni2022large}, SimLM~\cite{wang2023simlm}, BGE~\cite{xiao2024c}, and Instructor~\cite{su2023one}. 
For Mpnet, we use the strong version\footnote{\href{https://huggingface.co/sentence-transformers/all-mpnet-base-v2}{https://huggingface.co/sentence-transformers/all-mpnet-base-v2}} released by Sentence-Transformers~\cite{reimers-2019-sentence-bert}. 
Additionally, we evaluate the classic sparse retriever BM25~\cite{robertson2009probabilistic}. For retrievers that undergo instruction fine-tuning (see Table~\ref{tab:main_results}), we use the instructions provided in their respective papers for evaluation.

\vspace{2mm}

\input{table/main_results}

\noindent \textbf{Fine-tuned Baselines.} 
We follow the verbalizer-retriever approach from UDT-QA~\cite{ma2022open_2} to fine-tune a BERT-base model, serving as the UDT retriever. Since UDT focuses on homogeneous textual representations of heterogeneous data, we replace \(d_i\) with \(t_i\) from the data-text pairs \(\mathcal{D}\) during its training and evaluation, ensuring this model only interacts with the natural language corpus. This also means that in our experiments, we fine-tune the UDT-retriever baseline using exactly the same GPT-4o-mini synthesized data-text pairs \(\mathcal{D}\) as utilized by our UniHGKR.
For comparison, we also fine-tune a BERT-base model on the original CompMix-IR. Additionally, we finetune a DPR model using the UniK-QA method \cite{oguz2022unik}, serving as the UniK retriever. All fine-tuning uses the same positive and hard negative samples as UniHGKR. For baseline models lacking instruction-following capabilities, we input only the query across all retrieval scenarios to ensure optimal performance.

\subsection{Evaluation Metrics}

For retrieval scenario 1, we employ common metrics in the IR task: Hit@K (K=5,10,100) and MRR@K (Mean Reciprocal Rank, K=100) to evaluate model performance~\cite{zhao2024dense}.
More detailed descriptions are provided in Appendix~\ref{sec:app_metrics}.
For scenario 2, which uses type-specified instructions \(I_{\tau}\), where type \(\tau \in \mathcal{H}\), we introduce the metric Type-Hit (Type-Hit@100), indicate whether relevant evidence of the correct type is included in the top 100 retrieval result.

\subsection{Implementation Details.}
In our experiments, all contrastive training utilizes in-batch negatives across GPU devices. We utilize the maximum batch size that the GPU memory can fit and conduct all our training experiments on 8 A800-80GB GPUs.
In the training stage 3, each training sample has a group size of 16, which includes 1 positive sample and 15 hard negative samples. More detailed training settings can be found in Appendix~\ref{sec:training_setup}.

%% file: table/main_results.tex
\begin{table*}[!ht]
    \centering
    \begin{adjustbox}{max width=0.99\textwidth}
    \begin{tabular}{l|cc|cccc|cccc}
    \hline
    \rowcolor[rgb]{0.917, 0.929, 0.929}
      & & & \multicolumn{4}{c|}{ \cellcolor[RGB]{213, 211, 245} \textbf{Retrieval Scenario 1 (instruction \(\boldsymbol{I}_{\text{All}}\))}} &  \multicolumn{4}{c}{\cellcolor[RGB]{211, 240, 212} \textbf{Retrieval Scenario 2 (instruction \(\boldsymbol{I}_{\tau}\))}}  \\
      \hline
 \textbf{Method} & \textbf{Size}& \textbf{Ins}  &  \textbf{Hit@5} & \textbf{Hit@10}  & \textbf{Hit@100} & \textbf{MRR@100} &  \textbf{KG-Hit} & \textbf{Text-Hit} & \textbf{Table-Hit} & \textbf{Info-Hit}  \\ 
    \hline
BM25 & - & \textcolor{darkgray}{\ding{55}} & 11.51 & 17.40 & 52.39 & 8.54 & 24.20  & 34.55 &8.50 &19.79 \\

DPR & 109M & \textcolor{darkgray}{\ding{55}} & 24.89 & 36.32 & 78.76 & 17.51  &49.13 & 63.68 & 15.63 & 41.57  \\
Mpnet & 109M & \textcolor{darkgray}{\ding{55}} & 26.23 & 37.99 & 82.67 & 18.46 & 63.02  &61.11 & 18.96&52.1 \\
GTR-T5-base & 110M & \textcolor{darkgray}{\ding{55}} & 24.46 & 36.54 & 80.32 & 16.73 &57.78 & 59.8 & 22.87 & 46.09   \\
Contriever & 109M & \textcolor{darkgray}{\ding{55}} & 28.58 & 40.70 & 83.79 & 20.07 &62.26 & \underline{63.86} & 18.63 & 55.64 \\
SimLM & 109M & \textcolor{darkgray}{\ding{55}} & 25.11 & 37.08 & 80.61 & 17.68 &  59.59 & 59.01 & 17.69& 52.06 \\
Instructor-base & 110M & \cmarkgreenb & 24.86 & 36.22 & 81.55 & 17.80&  65.63 & 50.25&16.82 & 53.36 \\
Instructor-large & 336M & \cmarkgreenb & 25.98 & 36.87 & 81.51 & 18.54 &\underline{68.78} & 44.61 & 17.11 & 53.98  \\
BGE & 109M & \cmarkgreenb & 26.66 & 39.04 & 84.15 & 19.40  &68.42 & 57.96 & 22.58 & 56.58  \\
\hline
BERT-finetuned & 109M & \textcolor{darkgray}{\ding{55}} & 24.46 & 35.38 & 78.51 & 17.04   &57.63 & 54.67 & 17.55 & 48.41  \\

UDT-retriever & 109M & \textcolor{darkgray}{\ding{55}} & 24.96 & 35.49 & 76.52 & 18.24    &66.10 & 62.48 & 25.90  & \underline{57.05}   \\

UniK-retriever & 109M & \textcolor{darkgray}{\ding{55}} & \underline{30.68} & \underline{43.42} & \underline{85.20} & \underline{21.22}  &67.40 & 63.21 & \underline{26.74}& 56.04   \\

\hline
\textbf{UniHGKR-base} & 109M & \cmarkgreenb & \textbf{32.38} & \textbf{45.55} & \textbf{85.75} & \textbf{22.57} &\textbf{75.43} & \textbf{70.30}& \textbf{41.24} & \textbf{66.21}\\
\textcolor{darkred}{\ding{115}} Relative gain & & & \textcolor{darkred}{+5.54\%} & \textcolor{darkred}{+4.91\%} & \textcolor{darkred}{+0.65\%} & \textcolor{darkred}{+6.36\%} & \textcolor{darkred}{+9.67\%} & \textcolor{darkred}{+10.08\%} &  \textcolor{darkred}{+54.23\%} & \textcolor{darkred}{+16.06\%}  \\

\hline

\end{tabular}
\end{adjustbox}
\caption{the experimental results for the two retrieval scenarios on CompMix-IR. The relative gain is calculated based on the performance of UniHGKR-base compared to the best baseline, highlighted by \underline{underlines}.}
\label{tab:main_results}

\end{table*}


%% file: sections/result.tex
\section{Evaluation Results}
In this section, we focus on comparing and discussing the performance of UniHGKR with baselines on heterogeneous retrieval tasks and the application of UniHGKR models in the open-domain QA task. We also explore the robustness and zero-shot performance of UniHGKR in Appendix \ref{sec:retrieving_robustness}.


\subsection{Main Results}
\label{sec:main_results}
Table~\ref{tab:main_results} presents the retrieval performance of various models on the CompMix-IR test set. Our UniHGKR model outperforms all baselines in both scenarios, with a maximum relative improvement of 6.36\% in scenario 1 and 54.23\% in scenario 2, demonstrating its effectiveness in heterogeneous knowledge retrieval. Notably, powerful open-source retrievers like BGE (trained on over 200 million high-quality text pairs,~\cite{xiao2024c}) only achieve an MRR@100 below 20.0 in scenario 1 and a Table-Hit of 22.58 in scenario 2, highlighting the challenges of our constructed benchmark. 
Although the UDT-retriever shows significant improvement over its counterpart model, BERT-finetuned, in Scenario 2, the improvement in Scenario 1 is minimal. Also, it is clearly inferior to our UniHGKR-base, which was trained on the same synthetic data from GPT-4o-mini.
Moreover, UniK-retriever, fine-tuned the DPR model on CompMix-IR, performs well across several metrics but is suboptimal on structured data (like Table and Infobox) in scenario 2.
In contrast, our UniHGKR shows the greatest improvements on metrics where existing methods struggle, particularly in retrieving structured knowledge in scenario 2. This indicates that our three-stage training approach not only creates an effective representation space for heterogeneous knowledge retrieval but also excels at following diverse user instructions.

\subsection{Ablation Study}
\label{sec:ablation_study}
\input{table/ablation_table}

\input{table/unihgkr_llm}

In this subsection, we conduct ablation studies to examine the roles of different training stages and components in UniHGKR for heterogeneous knowledge retrieval. Table~\ref{tab:ablation_for_all_knowledge_source} presents the performance of various UniHGKR variants, obtained by removing specific components or a particular training stage.
Results show that removing any training stage or component leads to a significant drop in performance. For retrieval scenario 1, training stage 1 (Unified Embedding Self-Supervised Pretraining) is crucial, while for scenario 2, both rewritten (paraphrased) instructions and instruction-aware type-preferred loss $\mathcal{L}_{\text{preferred}}$ are key. Removing them will result in a performance drop of up to 13.42 points in the Table-Hit metric. Additionally, we present some extra ablation studies in Appendix~\ref{sec:additional_ablation}, such as exploring the role of different instructions in retrieving from specific sources and the performance gains of different training stages in an unsupervised setting.

\subsection{Extending UniHGKR to LLM Retrievers}
\label{sec:UniHGKR_llm}
Recent works, such as E5-mistral-7B \cite{wang2023improving} and LLARA \cite{li2023making}, have explored converting decoder-only LLMs into dense retrievers, leveraging their extensive pre-trained knowledge to achieve improvements on various IR tasks.
Our UniHGKR framework is plug-and-play and can seamlessly adapt to training LLM retrievers by adjusting the training objectives. To demonstrate this, we adapt UniHGKR framework to train our UniHGKR-7B retrievers based on the LLARA architecture. More adaptation details are in Appendix~\ref{sec:UniHGKR_7b_detail}.

Table~\ref{tab:unihgkr_llm} presents the evaluation results of UniHGKR-7B alongside other LLM-based baselines, including E5-mistral-7B, LLARA-passage (LLARA-pretrain fine-tuned on MS MARCO passage), and LLARA-finetuned (LLARA-pretrain fine-tuned on CompMix-IR). 
We can observe that our UniHGKR-7B significantly outperforms the LLM-based baselines and UniHGKR-base, achieving SOTA performance on all metrics across two scenarios. In particular, it achieves a 23.91\% relative improvement on the MRR@100 metric in Scenario 1 and reaches 49.57 on the Table-Hit in Scenario 2. These results further validate the effectiveness and scalability of our UniHGKR method, as well as the potential of LLMs as retrievers.

\subsection{Employing UniHGKR on QA systems}
\label{sec:qa_systems}
In this section, we explore the application of UniHGKR retrievers in open-domain QA systems over heterogeneous sources.
We select a popular task, ConvMix~\cite{christmann2022conversational}, which is a conversational format variant of the CompMix. This task is more challenging because it requires systems to consider both the current turn's question and the dialogue history. Baseline models such as QuReTeC~\cite{voskarides2020query}, CONVINSE~\cite{christmann2022conversational}, and EXPLAIGNN~\cite{christmann2023explainable}, along with their results, are sourced from the ConvMix leaderboard\footnote{\href{https://convinse.mpi-inf.mpg.de/}{https://convinse.mpi-inf.mpg.de/}}. Note that in the QA system experiment, we replace the entire retrieval component (e.g., CLOCQ+BM25) of the baseline with our UniHGKR model, not just BM25. The retrieval component of EXPLAIGNN and CONVINSE can be seen as a combination of coarse retrieval (CLOCQ) and re-ranking (BM25). All baseline methods and our UniHGKR use the same corpus to ensure a fair comparison. 
In the reasoning part after retrieval, we follow CONVINSE and use Fusion-in-Decoder (FiD)~\cite{izacard2021leveraging} as the reader. We input the top 100 relevant evidences returned by the retriever into the reader for inference. Then we evaluate the output of the reader as the performance of the QA system using the same metrics as baselines: P@1 (Precision at 1) and MRR. 

As shown in Table~\ref{tab:convmix_result}, by replacing retrievers with our UniHGKR models in baseline systems, we observe significant improvements in QA performance. 
Specifically, compared to CONVINSE, which uses the same reader FiD as we do, using UniHGKR-base as the retriever achieves an absolute improvement of up to 8.80 points in MRR, while UniHGKR-7B achieves an improvement of up to 13.60 points in MRR.
Compared to the current SOTA system, EXPLAIGNN, which uses a graph neural network (GNN) as a reader, our system surpasses it by up to 4.30 points in MRR and 5.90 points in P@1, setting a new SOTA performance for the ConvMix dataset.
These results further validate the effectiveness of UniHGKR and also indicate that the retrieval component is a significant factor limiting the performance of current open-domain QA systems on heterogeneous data.

\input{table/convmix_result}

%% file: table/ablation_table.tex
\begin{table*}[!h]
    \centering
    \begin{adjustbox}{max width=0.99\textwidth}
    \begin{tabular}{l|cccc|cccc}
    \hline
     \rowcolor[rgb]{0.917, 0.929, 0.929}
    & \multicolumn{4}{c|}{ \cellcolor[RGB]{213, 211, 245} \textbf{Retrieval Scenario 1 (instruction \(\boldsymbol{I}_{\text{All}}\))}} &  \multicolumn{4}{c}{\cellcolor[RGB]{211, 240, 212} \textbf{Retrieval Scenario 2 (instruction \(\boldsymbol{I}_{\tau}\))}} \\
     
    \textbf{Method} &  \textbf{Hit@5} & \textbf{Hit@10}  & \textbf{Hit@100} & \textbf{MRR@100} &  \textbf{KG-Hit} & \textbf{Text-Hit} & \textbf{Table-Hit} & \textbf{Info-Hit}     \\

    \hline
    \rule{0pt}{11pt}
\textbf{UniHGKR-base}      & \textbf{32.38} & \textbf{45.55} & \textbf{85.75}  & \textbf{22.57} & \textbf{75.43} & \textbf{70.30} & \textbf{41.24} & \textbf{66.21}      \\
\hline
\text{w/o training stage 1 (pretrain)}     & 29.78 & 42.80 & 84.88 & 21.54 & 72.97 & 68.60 & 34.01 & 60.06    \\

\text{w/o NL sentence during stage 1}     & 31.30 & 44.36 & 85.46 & 21.83 & 74.07 & 69.01 & 40.13 & 	65.16    \\

\text{w/o training stage 2 (alignment)}     & 31.41 & 45.02 & 85.14 & 21.92 & 74.75 & 70.01 & 37.99 & 66.04    \\

\hline
\rule{0pt}{13pt}
\text{In the training stage 3 (finetune)}     &  &  &  &  &  &  &  &      \\

\text{\ \ w/o type-preferred loss \(\mathcal{L}_{\text{preferred}}\)} & 31.77 & 44.97 & 85.24 & 22.27 & 73.26 & 69.90 & 33.86 & 61.22 \\

\text{\ \ w/o instructions and \(\mathcal{L}_{\text{preferred}}\)}  & 31.98 & 44.39 & 85.24 & 22.18 & 68.20 & 65.59 & 29.20  &  58.65  \\

\text{\ \ w/o rewritten instructions }  & 31.11 & 44.14 & 84.48 & 21.86 & 67.66 & 65.16 & 27.82  &  57.96  \\

\hline

\end{tabular}
\end{adjustbox}
\caption{The results of the ablation study for the UniHGKR-base. We use blue color to indicate the largest decrease.}
\label{tab:ablation_for_all_knowledge_source}

\end{table*}

%% file: table/unihgkr_llm.tex
\begin{table*}[!h]
    \centering
    \begin{adjustbox}{max width=0.93\textwidth}
    \begin{tabular}{l|cccc|cccc}
    \hline
     \rowcolor[rgb]{0.917, 0.929, 0.929}
    & \multicolumn{4}{c|}{ \cellcolor[RGB]{213, 211, 245} \textbf{Retrieval Scenario 1 (instruction \(\boldsymbol{I}_{\text{All}}\))}} &  \multicolumn{4}{c}{\cellcolor[RGB]{211, 240, 212} \textbf{Retrieval Scenario 2 (instruction \(\boldsymbol{I}_{\tau}\))}} \\
     
  \textbf{Method}   &  \textbf{Hit@5} & \textbf{Hit@10}  & \textbf{Hit@100} & \textbf{MRR@100}   &  \textbf{KG-Hit} & \textbf{Text-Hit} & \textbf{Table-Hit} & \textbf{Info-Hit} \\

    \hline
UniHGKR-base  & 32.38 & 45.55 & 85.75 & 22.57 & 75.43 & 70.30 & 41.24 & 66.21\\
\hline
\text{E5-mistral-7B}      & 31.3 & 43.49 & 83.36 & 22.97  & 69.03 & 41.46 & 33.03 & 62.92  \\

\text{LLARA-passage}     & 37.45 & 51.59 & 86.61 & 26.11 & 68.23 &  \underline{70.48} &\underline{37.88} & 60.64  \\

\text{LLARA-finetuned}     & \underline{42.19} & \underline{55.35} & \underline{87.81} & \underline{30.83}  &  \underline{74.38} & 69.86 & 36.40 & \underline{64.40}  \\

\hline
\textbf{UniHGKR-7B}     & \textbf{49.78} & \textbf{59.23} & \textbf{88.21} & \textbf{38.20}  &\textbf{81.80} & \textbf{76.05} & \textbf{49.57} &\textbf{73.88} \\
\textcolor{darkred}{\ding{115}}Relative gain & \textcolor{darkred}{+17.99\%} & \textcolor{darkred}{+7.01\%} &  \textcolor{darkred}{+0.46\%} & \textcolor{darkred}{+23.91\%}  & \textcolor{darkred}{+9.98\%}   & \textcolor{darkred}{+7.90\%}  & \textcolor{darkred}{+30.86\%}  & \textcolor{darkred}{+14.72\%}  \\

\hline

\end{tabular}
\end{adjustbox}
\caption{Retrieval performances of UniHGKR-7B and LLM-based retrievers baselines. The relative gain is calculated based on the performance of UniHGKR-7B compared to the best baseline, highlighted by \underline{underlines}.}
\label{tab:unihgkr_llm}
\vspace{-2mm}
\end{table*}

%% file: table/convmix_result.tex
\begin{table}[t]
    \centering
    \begin{adjustbox}{max width=0.49\textwidth}
    \begin{tabular}{l|cc|ccc}
    \hline
     \rowcolor[rgb]{0.917, 0.929, 0.929} \textbf{Methods} & \textbf{Retriever} & \textbf{Reader} & \textbf{P@1} & \textbf{MRR}&  \\
    \hline
    
    \text{BM25+FiD} & BM25 & FiD & 	25.3 & 27.5  \\
    \text{QuReTeC} & QuReTeC & FiD & 28.2 & 28.9  \\
    \text{CONVINSE} & CLOCQ+BM25 & FiD & 34.3 & 37.8  \\
    \text{EXPLAIGNN} & CLOCQ+BM25 & GNN & \underline{40.6} & \underline{47.1}  \\
    \hline
    \textbf{Ours} & \textbf{UniHGKR-base} & FiD & 42.4 & 46.6   \\
       \textcolor{blue}{\ding{115}}Abs. gain && & \textcolor{blue}{+8.10} &  \textcolor{blue}{+8.80}  \\
       
   & \textbf{UniHGKR-7B} & FiD & \textbf{46.5} & \textbf{51.4}   \\
   \textcolor{blue}{\ding{115}}Abs. gain && & \textcolor{blue}{+12.20} &  \textcolor{blue}{+13.60}  \\
   \hline 
   \textcolor{darkred}{\ding{115}}SOTA gain && & \textcolor{darkred}{\textbf{+5.90}} &  \textcolor{darkred}{\textbf{+4.30}}  \\
\hline
\end{tabular}
\end{adjustbox}
    \caption{The QA performance of systems using the UniHGKR retriever and baselines on the ConvMix dataset. `Abs. gain' represents the absolute improvement brought by the retriever under the same Reader setting (compared to CONVINSE). `SOTA gain' indicates the absolute improvement over the previous SOTA system.}
\label{tab:convmix_result}
\vspace{-3mm}
\end{table}


%% file: sections/conclusion.tex
\section{Conclusion}
In this paper, we introduced UniHGKR, an instruction-aware unified heterogeneous knowledge retriever. First, we constructed CompMix-IR, the first heterogeneous information retrieval task dataset containing a corpus of over 10 million entries across four heterogeneous data types. Then, we defined two different heterogeneous information retrieval scenarios to meet the diverse retrieval needs of real-world users. We designed the UniHGKR framework with three training stages. Our experiments showed that UniHGKR achieved state-of-the-art performance on CompMix-IR benchmarks, both with the 110M BERT-based retriever and the 7B LLM-based retriever. Applying our UniHGKR retrievers can significantly enhance the performance of heterogeneous QA systems, achieving new SOTA results on the ConvMix dataset.

%% file: sections/limitations.tex

\section{Limitations}
In our study, the CompMix-IR dataset is primarily sourced from Wikidata knowledge graphs and Wikipedia, including infoboxes, tables, and text, but it is limited to five domains: books, movies, music, television series, and football. This may restrict the model’s generalization capabilities. Additionally, while UniHGKR incorporates diverse user instructions, it does not cover all scenarios in heterogeneous information retrieval. For instance, users might want to instruct the retriever to return a combination of evidence from multiple knowledge sources, such as text and tables, or a mix of KG triples, tables, and text, as noted in \cite{christmann2022conversational}. Exploring these user-defined combinations remains an area for future work. In addition, more modalities such as image, audio and interleaved image and text~\cite{xu2024lateralizationlorainterleavedinstruction} can be considered and incorporated in the retrieving process of UniHGKR in future. We will open-source our instruction set, CompMix-IR corpus, and UniHGKR model and code, encouraging the community to contribute more retrieval tasks with large-scale human-written instructions \cite{visionFlan} to assess whether broader instruction coverage enhances performance.


%% file: sections/appendix.tex
\section{CompMix-IR Example}
\label{sec:appendix_compmix_ir}

\subsection{Heterogeneous Evidence Examples}
Table~\ref{tab:compmix_ir_examples_evi} provides linearized heterogeneous examples of evidence for the four types of knowledge. Table~\ref{tab:compmix_ir_examples_with_annotation} provides examples of evidence with full annotation information.

\begin{table}[h!]
\small
\begin{tcolorbox}

\textcolor{red}{\textbf{Text evidences:}}
\\
1. Cousteau (band), Reboot In 2016 it was announced that Liam McKahey and Davey Ray Moor were returning as CousteauX and were back in the recording studio preparing new music.\\
2. Cousteau (band), To honour the new era the band placed an X at the end of their name.\\
3. Cousteau (band), Cousteaux is another popular French family name.\\
\\

\textcolor{red}{\textbf{KG evidences:}}
\\
1. Maverick, cast member, Robert Colbert\\
2. Maverick, original language of film or TV show, English \\
3. Maverick, cast member, Roxane Berard, name of the character role, `Comtesse de Barot', name of the character role, `Comtesse Lizette de La Fontaine', name of the character role, `Felice de Lassignac', name of the character role, `Danielle de Lisle'\\
\\

\textcolor{red}{\textbf{Table evidences:}}
\\
1. Stefanie Powers, Year is 1975, Title is Gone with the West, Role is Little Moon, Notes is Alternate title: Little Moon and Jud McGraw\\
2. Stefanie Powers, Year is 1975, Title is It Seemed Like a Good Idea at the Time, Role is Georgia Price, Notes is.\\
3. Stefanie Powers, Year is 1976, Title is Invisible Strangler, Role is Candy Barrett, Notes is Alternate titles: The Astral Factor , The Astral Fiend\\
\\

\textcolor{red}{\textbf{Infobox evidences:}}
\\
1. When Harry Met Sally..., When Harry Met Sally…, Directed by, Rob Reiner \\
2. When Harry Met Sally..., When Harry Met Sally…, Written by, Nora Ephron \\
3. When Harry Met Sally..., When Harry Met Sally…, Produced by, Rob Reiner Andrew Scheinman\\

\end{tcolorbox}
\caption{Evidence examples from the CompMix-IR corpus.}
\label{tab:compmix_ir_examples_evi}
\end{table}

\begin{table}[h!]
\small
\begin{tcolorbox}

\textcolor{red}{\textbf{Text evidence:}}
\\
\{ "linearized evidence text": "Museum of Modern Art, Its first successful loan exhibition was in November 1929, displaying paintings by Van Gogh, Gauguin, Cézanne, and Seurat.",\\
"wikidata entities": [ \{ "id": "Q34013", "label": "Georges Seurat" \}, \{ "id": "Q17437796", "label": "featured article" \}, ...],\\
"disambiguations": [ [ "1929", "1929-01-01T00:00:00Z" ], [ "painting", "Q11629" ], [ "Van Gogh", "Q17437796" ], … ],\\
"retrieved for entity": \{ "id": "Q188740", "label": "Museum of Modern Art" \},\\
"source": "text" \}
\\

\textcolor{red}{\textbf{KG evidence:}}
\\
\{"linearized evidence text": "Transformers, genre, action film",\\ 
"wikidata entities": [ \{ "id": "Q171453", "label": "Transformers" \}, \{ "id": "Q188473", "label": "action film" \} ],\\
"disambiguations": [ [ "Transformers", "Q171453" ],…],\\
"source": "kg"\}\\
\\
\textcolor{red}{\textbf{Table evidence:}}\\
\{ "linearized evidence text": "Surrey Scorchers, Season is 2008–09, Division is BBL, Tier is I, Regular Season is 4th, Post-Season is 33, Trophy is 21, Cup is 12, Head Coach is 42",\\
"wikidata entities": [ \{ "id": "Q3645013", "label": "2008–09 British Basketball League season" \}, \{ "id": "Q269597", "label": "Surrey Scorchers" \}, \{ "id": "Q23276", "label": "Surrey"\}, ... ],\\
"disambiguations": [ [ "2008", "2008-01-01T00:00:00Z" ], [ "2008–09", "Q3645013" ], ...],\\
"retrieved for entity": \{ "id": "Q269597" \},\\
"source": "table"\}\\
\\
\textcolor{red}{\textbf{Infobox evidence:}}
\\
\{ "linearized evidence text": "Older (Royseven song), from the album '' The Art of Insincerity, Genre, Rock",\\
"wikidata entities": [ \{ "id": "Q7375238", "label": "Royseven" \}, \{ "id": "Q188451", "label": "music genre" \}, \{ "id": "Q7714454", "label": "The Art of Insincerity" \}, ... ],\\
"disambiguations": [ [ "The Art of Insincerity", "Q7714454" ], [ "Royseven", "Q7375238" ], [ "Genre", "Q188451" ], ... ],\\
"retrieved for entity": \{ "id": "Q7085553" \},\\
"source": "info"
\}

\end{tcolorbox}
\caption{Evidence examples with full annotation information.}
\label{tab:compmix_ir_examples_with_annotation}
\end{table}

\subsection{CompMix-IR QA Examples}

We present some question-answer examples from the CompMix-IR dataset in Table~\ref{tab:compmix_ir_examples_qa}, while Table~\ref{tab:compmix_ir_examples_qa_with_annotation} provides a QA example with full annotation information. Table ~\ref{tab:compmix_qa_statistics} shows the statistics of the CompMix-IR QA set.

\begin{table}[ht]
\small
\begin{tcolorbox}

\textbf{Question 1:}
\\
Which is the initial book of the book series Divergent?\\
\\
\textcolor{blue}{\textbf{Answer 1:}\\Divergent (novel).}\\
\rule{\linewidth}{0.2mm}
\textbf{Question 2:}
\\
Who is the author of the book Divergent (novel)?\\
\\
\textcolor{blue}{\textbf{Answer 2:}\\
Veronica Roth.}\\
\rule{\linewidth}{0.2mm}
\textbf{Question 3:}
\\
What is the date of birth of the Divergent's author Veronica Roth?\\
\\
\textcolor{blue}{\textbf{Answer 3:}\\
19 August 1988}\\
\end{tcolorbox}
\caption{QA examples from the CompMix-IR dataset.}
\label{tab:compmix_ir_examples_qa}
\end{table}

\begin{table}[ht]
\small
\begin{tcolorbox}

\textbf{Question:}
\\
Who was the voice actor for Meg Griffin in Family Guy?\\
\\
\textbf{Answer:}\\Mila Kunis\\

\textbf{Annotation information:}
\{"question id": "5136",\\
"question": "Who was the voice actor for Meg Griffin in Family Guy?",\\
"domain": "tvseries",\\
"entities": [\{"id": "Q908772", "label": "Meg Griffin"\}, \{"id": "Q5930", "label": "Family Guy"\}],\\
"answers": [\{"id": "Q37628", "label": "Mila Kunis"\}],\\
"answer text": "Mila Kunis",\\
"answer src": "kg"\}

\end{tcolorbox}
\caption{A QA example with full annotation information.}
\label{tab:compmix_ir_examples_qa_with_annotation}
\end{table}

\begin{table}[ht]
\centering\resizebox{0.39\textwidth}{!}
{
\begin{tabular}{l|c|rc} 

\hline
\multicolumn{2}{c|}{\textbf{Dataset}}& \multicolumn{2}{c}{\textbf{Question word (\%)}} \\ \hline

Train set& 4,966 & What& 39.28            \\
Dev set& 1,680 & Who& 29.69          \\
Test set& 2,764& Which& 16.90                \\
Total& 9,410 & How& 5.48              \\ 

\cline{1-2}
\multicolumn{2}{c|}{\textbf{Avg. length}} & When& 5.13 \\ 
\cline{1-2}
Question                & 9.19 & Where& 3.32                     \\
Answer   & 2.17               & Other & 0.20 \\ 

\hline
\end{tabular}
}
\caption{Question answering Statistics of CompMix-IR. `Avg. length' refers to the average number of words.}
\label{tab:compmix_qa_statistics}
\end{table}

\section{Prompt Example}
Table~\ref{tab:prompt_examples} shows a prompt example we use in constructing Data-text Pairs, with the help of GPT-4o-mini.
\label{sec:prompt_tempplate}

\begin{table}[h]
\small
\begin{tcolorbox}


\textbf{Prompt template:}

Evidence data is a triple from the wikidata
knowledge graph, representing a factual piece
of information. The components of the triple
are separated by ', ' and represent the head
entity, the relation, and the tail entity,
respectively. I hope you understand the content
of evidence data, and then use grammatically
correct natural language sentence to describe
the content in evidence data. Here has some
demonstrations:
\textbf{<Demonstrations>}

\end{tcolorbox}
\vspace{-2mm}
\caption{The prompt example used in KG triples.}
\label{tab:prompt_examples}
\end{table}

\section{Detailed descriptions of  Metrics}
\label{sec:app_metrics}
In our study, we use the following metrics to measure retrieval performance:
\begin{itemize}
    \item Hit@K, also known as Top-k Accuracy~\cite{karpukhin2020dense}, measures the proportion of queries for which the top-k retrieved evidence contains the correct answers. This is a key metric for retrievers in the RAG framework.
    \item Mean Reciprocal Rank (MRR)~\cite{zhao2024dense} computes the average of the reciprocal ranks of the first relevant evidence retrieved across a set of queries.
\end{itemize}


\section{Training setup}
\label{sec:training_setup}
In this section, we detail the detailed training settings for training UniHGKR-base and UniHGKR-7B. In training phase 3, a larger number of instruction-unfollowing negative samples could potentially harm the performance of the retriever in retrieval scenario 1. Therefore, in our training, we set a probability of 0.005 to add one instruction-unfollowing negative sample in the training samples of retrieval scenario 2.

\subsection{UniHGKR-base Training setup}
During training phase 1, we initialize model parameters from BERT-base~\cite{devlin-etal-2019-bert} weights. The learning rate is set to \(1 \times 10^{-5}\). Training is conducted for one epoch with a batch size of 32 per device. In training phase 2, the learning rate increases to \(2 \times 10^{-5}\). Training also spans one epoch, but the batch size per device increases to 96. In-batch negative samples can be used across devices, increasing the diversity and number of negative samples used during training. In the subsequent training phase 3, the learning rate remains \(2 \times 10^{-5}\), but the training duration is extended to 5 epochs. The batch size per device is reduced back to 32 to accommodate a larger hard negative sample group, with a size of 15. In training phases 2 and 3, the temperature parameter is set to 0.02, and both phases use FP16 precision mode to enhance computational efficiency and conserve memory.

\subsection{UniHGKR-7B Training setup}
In the initial training phase (stages 1 and 2), we initialize model parameters from LLARA-pretrain~\cite{li2023making} weights. The learning rate is set to \(1 \times 10^{-5}\), with a batch size of 384 per device, for one epoch. In these stages, we use the full parameter training method.
In the third training phase, we increase the learning rate to \(2 \times 10^{-4}\) and reduce the batch size per device to 64 to accommodate a larger negative sample group size of 7. Training is conducted for one epoch. During this phase, we introduce parameter-efficient training method LoRA~\cite{hu2022lora} with a rank of 64 and an alpha value of 16. The dropout rate for LoRa is set to 0.1 to prevent overfitting. Similar to UniHGKR-base, we enable in-batch negative sampling across devices to increase the diversity and number of negative samples during training.

\begin{figure}[h]
\centering 
     \includegraphics[width=0.45\textwidth]{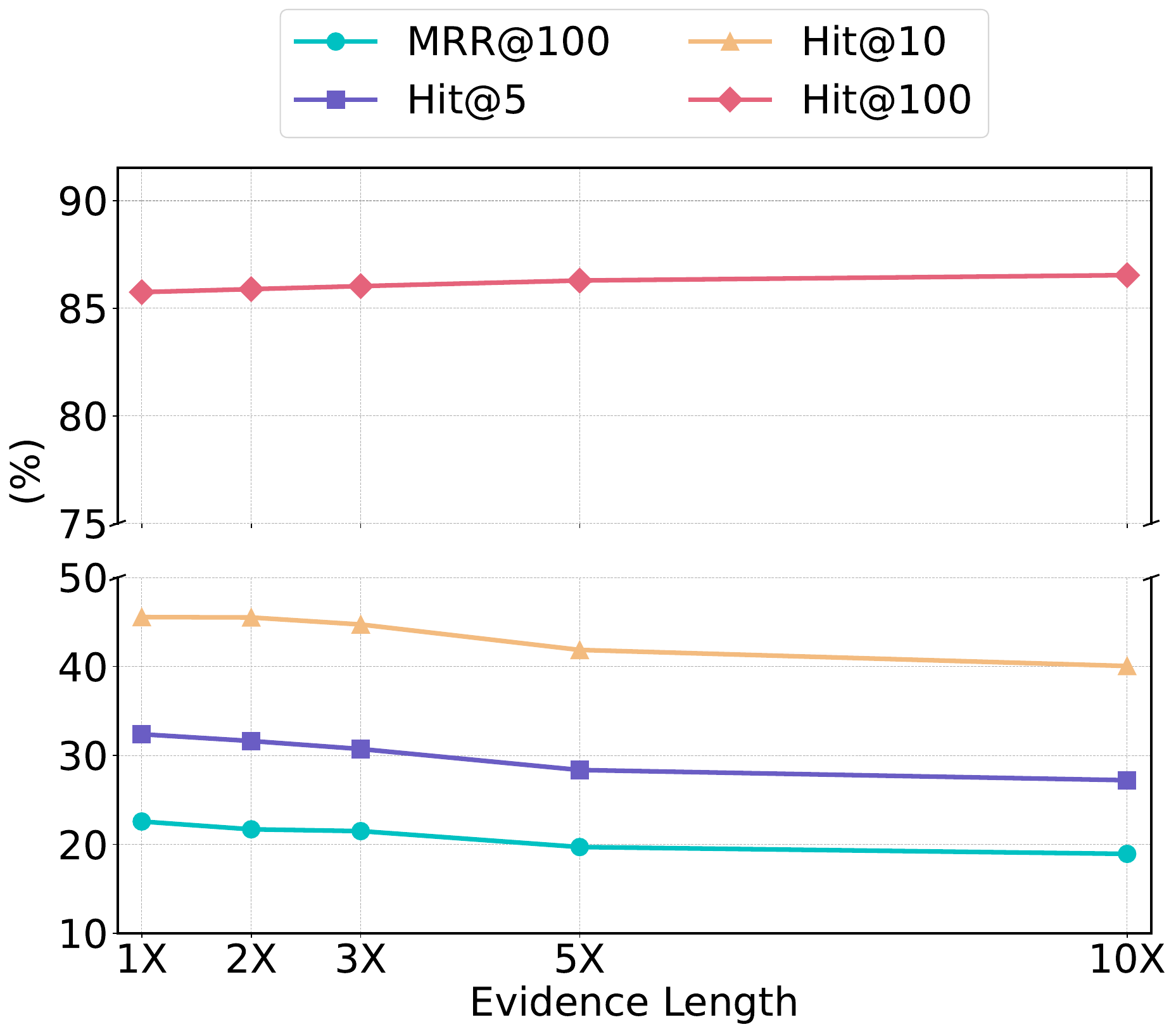}
     \caption{The performance of UniHGKR-base in retrieval Scenario 1 with longer evidences. Here, 10X indicates that the average length of the evidence in the corpus is 10 times the original (1X), and so on.}
     \label{fig:evidence_length}
\end{figure}

\section{Retrieving Robustness of UniHGKR}
\label{sec:retrieving_robustness}
In this section, we evaluate the performance of the UniHGKR-base model on longer evidence corpora, as well as its zero-shot generalization capabilities.

\vspace{2mm}

\begin{figure}[h]
\centering 
     \includegraphics[width=0.49\textwidth]{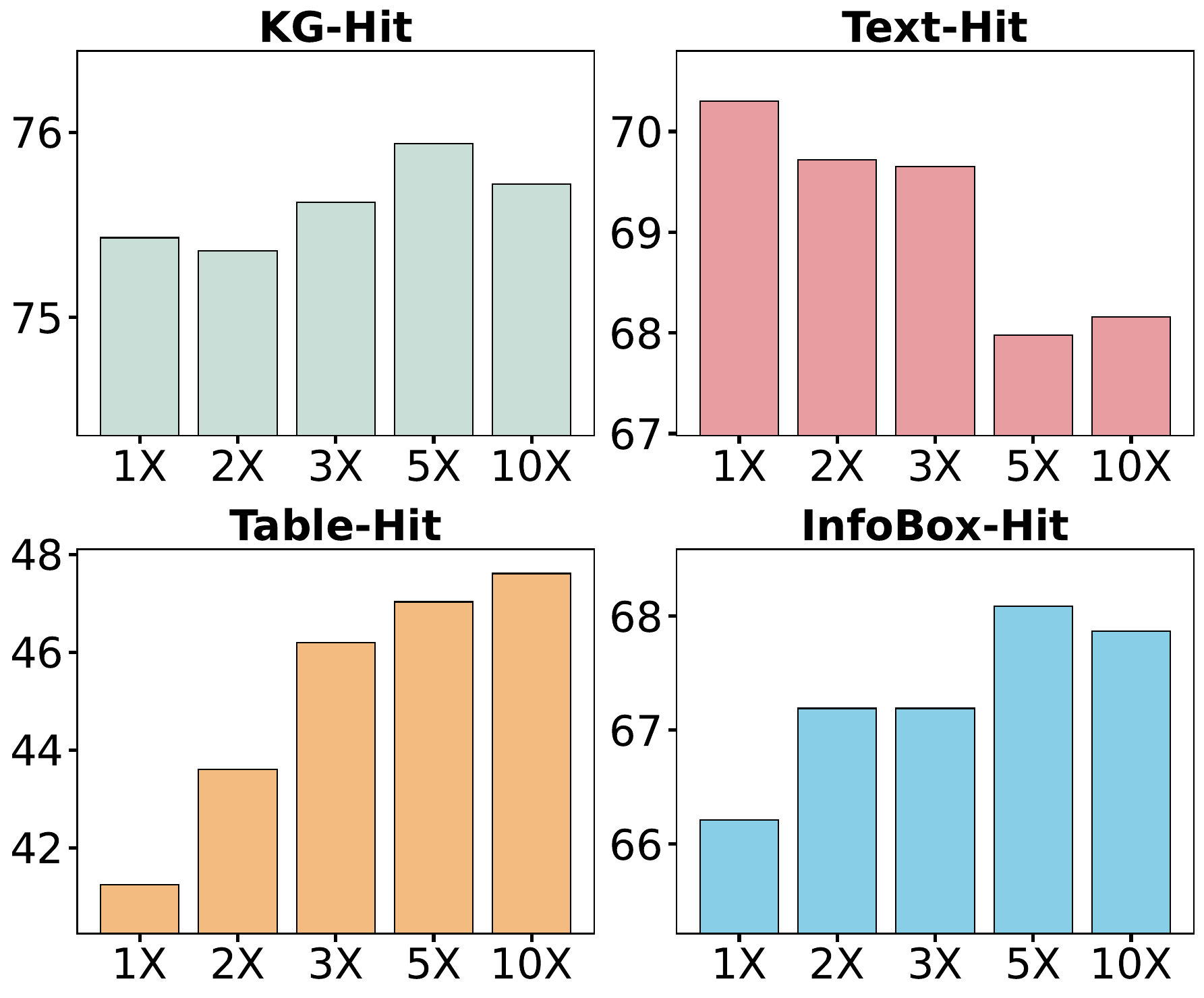}
     \caption{The performance of UniHGKR-base in retrieval Scenario 2 with longer evidences.}
     \label{fig:4_types_bar_evi_lens}
\end{figure}

\noindent \textbf{Robustness for Evidence Length.} The robustness of retrievers to varying evidence lengths is crucial, as dense retrievers encounter varying inputs lengths in real-world applications. By increasing the segmentation size of the evidence during the construction of the CompMix-IR corpus, we create several corpus variants,the average length of whose evidence is 2 to 10 times that of the original version. We then evaluate UniHGKR-base, which is trained on the original CompMix-IR corpus, for its retrieval performance on these longer corpus variants, as shown in Figures~\ref{fig:evidence_length} and \ref{fig:4_types_bar_evi_lens}. From Figure~\ref{fig:evidence_length}, we can see that our UniHGKR-base model shows good robustness with respect to evidence length in retrieval scenario 1. Its performance on metrics like MRR@100 and Hit@5 shows only a slight decline as the evidence length increases, while the Hit@100 metric even shows improvement. This may be because longer evidence can include more information within the fixed number (top-100) evidences, consistent with the findings in~\cite{jiang2024longrag}. On the other hand, Figure \ref{fig:4_types_bar_evi_lens} shows the retrieval scenario 2 performance of retrieving specified knowledge types on longer evidence. An interesting finding is that the performance of UniHGKR-base in retrieving longer structured data evidence does not decline. Instead, it experiences varying degrees of improvement, most notably on the Table-Hit, where it increases by more than 6 points. This may be because longer evidence can prevent long structured data, such as tables with many rows and columns, from being fragmented into multiple parts, thus avoiding semantic loss.

\noindent \textbf{Zero-Shot Performance on BEIR.} An advantage of instruction-aware universal heterogeneous knowledge retrievers is their enhanced ability to generalize to unseen domains with various types of evidence candidates. To validate this, we evaluate the zero-shot retrieval performance of UniHGKR-base on the popular IR benchmark BEIR~\cite{thakur2021beir}. This benchmark includes domains not encountered during UniHGKR's training, such as Bio-Medical and Finance. Following standard setting~\cite{xiao2022retromae,liu2023retromae}, we fine-tune the pre-trained model with MS MARCO~\cite{nguyen2017ms} and evaluate zero-shot transferability on the other 12 datasets. Following~\cite{thakur2021beir}, for BEIR,  we use NDCG@10 as our primary metric on BEIR. Results for baselines like BERT, SimCSE~\cite{gao-etal-2021-simcse}, and DiffCS~\cite{chuang2022diffcse} are taken from \cite{xiao2022retromae}. As shown in Table~\ref{tab:beir}, our UniHGKR model demonstrates strong zero-shot generalization capabilities. It outperforms baselines on the unseen domain IR datasets, such as the Bio-Medical domain TREC-COVID~\cite{voorhees2021trec} and the Finance domain FiQA-2018~\cite{maia201818}, while maintaining a clear advantage on the familiar task: Wikipedia Entity-Retrieval dataset DBPedia~\cite{hasibi2017dbpedia}. Additionally, UniHGKR-base also demonstrates clear advantages over the baselines on pure natural language text QA information retrieval datasets, such as NQ~\cite{kwiatkowski2019natural} and HotpotQA~\cite{yang2018hotpotqa}. We believe this is because, through our training stages 1 and 2, the model has learned better capabilities to capture the essence of semantic information, which is beneficial for a wide range of retrieval tasks.

\input{table/beir}

\section{Additional Ablation Studies}
\label{sec:additional_ablation}

\subsection{Experiments under the Unsupervised Setting}
We conduct experiments under the unsupervised setting (i.e., after training in Stage 1 and Stage 2) in retrieval scenario 1, and the results are shown in Table~\ref{tab:unsupervised_retrieval_result}. From these results, we can clearly observe the performance gains brought by each stage to the model's retrieval capabilities. Overall, the alignment training in Stage 2 provides more significant gains compared to the pretraining in Stage 1. After training in Stage 2, the unsupervised model achieves a respectable 73.52 in Hit@100.

\begin{table}[ht]
    \centering
    \begin{adjustbox}{max width=0.49\textwidth}
    \begin{tabular}{l|cccc}
    \hline
    \rowcolor[rgb]{0.917, 0.929, 0.929} \textbf{Method} & \textbf{Hit@5} & \textbf{Hit@10} & \textbf{Hit@100} & \textbf{MRR@100} \\
    \hline
    \text{Bert-base-uncased} & 6.55 & 10.93 & 37.19 & 5.04 \\
    \hline
    \text{After Stage 1} & 9.26 & 14.47 & 49.78 & 6.76 \\
    \textcolor{blue}{\ding{115}}Abs. gain & \textcolor{blue}{+2.71} & \textcolor{blue}{+3.54} & \textcolor{blue}{+12.59} & \textcolor{blue}{+1.72} \\
    \hline
    \text{After Stage 2} & 16.03 & 25.54 & 73.52 & 12.10 \\
    \textcolor{blue}{\ding{115}}Abs. gain & \textcolor{blue}{+6.77} & \textcolor{blue}{+11.07} & \textcolor{blue}{+23.74} & \textcolor{blue}{+5.34} \\
    \hline
    \textbf{After Stage 3*} & \textbf{32.38} & \textbf{45.55} & \textbf{85.75} & \textbf{22.57} \\
    \textcolor{blue}{\ding{115}}Abs. gain & \textcolor{blue}{+16.35} & \textcolor{blue}{+20.01} & \textcolor{blue}{+12.23} & \textcolor{blue}{+10.47} \\
    \hline
    \end{tabular}
    \end{adjustbox}
    \caption{The performance on retrieval scenario 1 after different training stages. Among them, `After Stage 1' and `After Stage 2' can be regarded as the performance in the unsupervised setting. `After Stage 3*' represents our UniHGKR-base model. `Abs. gain' represents the absolute improvement in performance after each training stage.}
    \label{tab:unsupervised_retrieval_result}
    \vspace{-3mm}
\end{table}

\subsection{The Impact of Instructions for Retrieving from Specific Sources.}
We added experiments on retrieving from specific sources in the \( I_{\text{All}} \) setting. Based on this, we can compare and observe the improvement in performance when using the instruction \( I_{\tau} \), which specifies the retrieval source, in retrieval scenario 2.
In Table~\ref{tab:retrieval_specific_sources}, we can clearly see that when retrieving specific types of knowledge, our UniHGKR model shows a significant improvement when using the instruction \( I_{\tau} \) (where \( \tau \in \mathcal{H} = \text{Text, Info, Table, KG} \)) compared to using the instruction \( I_{\text{All}} \). This is particularly the case for table and infobox-type knowledge. This result indicates that our proposed type-preferred loss (\( \mathcal{L}_{\text{preferred}} \)) can help the model distinguish data types and capture their differences for flattened inputs with the help of instructions.

\begin{table*}[h]
    \centering
    \begin{adjustbox}{max width=0.7\textwidth}
    \begin{tabular}{l|c|c|c|c|c}
    \hline
    \rowcolor[rgb]{0.917, 0.929, 0.929} \textbf{Method} & \textbf{Instructions} & \textbf{KG-Hit} & \textbf{Text-Hit} & \textbf{Table-Hit} & \textbf{Info-Hit} \\
    \hline
    UniHGKR-base & \( I_{\text{All}} \) & 68.60 & 65.70 & 28.76 & 57.34 \\
    UniHGKR-base & \( I_{\tau} \) & 75.43 & 70.30 & 41.24 & 66.21 \\
    \hline
    \textbf{Abs. gain} & & \textbf{+6.83\%} & \textbf{+4.60\%} & \textbf{+12.48\%} & \textbf{+8.87\%} \\
    \hline
    \end{tabular}
    \end{adjustbox}
    \caption{Performance of retrieving specific knowledge types with different instructions in retrieval scenario 2. `Abs. gain' refers to the performance improvement brought by using instruction \( I_{\tau} \) compared to \( I_{\text{All}} \).}
    \label{tab:retrieval_specific_sources}
\end{table*}

\begin{table*}[h]
    \centering
    \begin{adjustbox}{max width=0.9\textwidth}
    \begin{tabular}{l|c|c|c|c}
    \hline
    \rowcolor[rgb]{0.917, 0.929, 0.929} \textbf{Model} & \textbf{Size} & \textbf{Vector Dim.} & \textbf{Avg. Embed Time (100 evd)} & \textbf{Avg. Retrieve Time (100 ques)} \\
    \hline
    UniHGKR-Base & 109M & 768 & 0.46 s & 4.35 s \\
    UniHGKR-7B & 7B & 4096 & 1.54 s & 53.29 s \\
    \hline
    \end{tabular}
    \end{adjustbox}
    \caption{Time efficiency comparison between UniHGKR-Base and UniHGKR-7B. The experiment was conducted on a single V100-32G GPU on the CompMix-IR. The data are the average values of three runs of the experiment for 100 pieces of evidence or 100 questions.}
    \label{tab:time_efficiency}
\end{table*}

\subsection{Efficiency of the Proposed Models}

For retrieval tasks, efficiency is as important as accuracy. The time cost of retrieval tasks lies in two parts: (1) Embedding, (2) Retrieving. The factor affecting the first part `Embedding' is the parameter scale of the dense embedder. So, the parameter scales of the baselines and UniHGKR models are shown in Table~\ref{tab:main_results} and Table~\ref{tab:unihgkr_llm}. The efficiency of the second part `Retrieving' is affected by the dimension of the vector generated by the retriever. We added an experiment to show the time efficiency difference between the UniHGKR-Base model and the UniHGKR-7B model, as shown in Table~\ref{tab:time_efficiency}. The embedding and retrieving average time costs for UniHGKR-7B are 3.35 and 12.25 times longer than those for UniHGKR-Base, respectively. Note that during retrieval, we did not use fast vector retrieval libraries such as Faiss~\cite{johnson2019billion} but instead performed a naive KNN~\cite{steinbach2009knn} computation.

\section{Detailed Description of UniHGKR-7B Adaptation}
\label{sec:UniHGKR_7b_detail}

In our UniHGKR-7B training, we initialize the model weights from the LLaRA-pretrain. LLARA-pretrain model initializes its parameters from LLaMA-2-7B-base~\cite{touvron2023llama}. The output vector of the last token of the model input sequence \(S\), a special token \(\langle\backslash s\rangle\), is used as the embedding representation \(r\) of the input sequence:
\[r \leftarrow \text{LLaMA}(S)[\langle\backslash s\rangle].\]
They then apply their proposed Embedding Based AutoEncoder (EBAE) and Embedding Based AutoRegressive (EBAR) techniques for post-training adaptation for dense retrieval. EBAE reconstructs the tokens of the input sentence using $r$, while EBAR predicts the tokens of the next sentence based on $r$.

In Stages 1 and 2 of our UniHGKR-7B training, our input sequence \(S\) is the linearized structured data \(d_i\). We adapt EBAE to reconstruct \(d_i\) and EBAR to predict the corresponding natural language sentence \(t_i\). Here, \(\langle d_i, t_i \rangle \) is from the Data-Text Pairs \(\mathcal{D}\). This process essentially implements Stages 1 and 2 of our UniHGKR training framework: establishing an effective representation space for heterogeneous knowledge. For task fine-tuning (Stage 3), we use the same training methods as the UniHGKR-base models (BERT-based), including the instruction set and positive/negative sampling strategies (see Section~\ref{sec:UniHGKR_train}).


%% file: table/beir.tex
\begin{table}[h!]
    \centering
    \begin{adjustbox}{max width=0.49\textwidth}
    \begin{tabular}{l|cccc}
    \hline    \rowcolor[rgb]{0.917, 0.929, 0.929}
    \textbf{Datasets} & \textbf{BERT} & \textbf{SimCSE} & \textbf{DiffCSE} & \textbf{UniHGKR} \\
    \hline
    \text{TREC-COVID} & \underline{0.615} & 0.460 & 0.492 & \textbf{0.650} \\
    \text{NFCorpus} & \underline{0.260} & \underline{0.260} & 0.259 & \textbf{0.279} \\
    \text{NQ} & \underline{0.467} & 0.435 & 0.412 & \textbf{0.490} \\
    \text{HotpotQA} & 0.488 & \underline{0.502} & 0.499 & \textbf{0.525} \\
    \text{FiQA-2018} & \underline{0.252} & 0.250 & 0.229 & \textbf{0.261} \\
    \text{ArguAna} & 0.265 & \underline{0.413} & \textbf{0.468} & 0.400 \\
    \text{Touche-2020} & \textbf{0.259} & 0.159 & 0.168 & \underline{0.202} \\
    \text{DBPedia} & \underline{0.314} & \underline{0.314} & 0.303 & \textbf{0.334} \\
    \text{SCIDOCS} & 0.113 & 0.124 & \underline{0.125} & \textbf{0.133} \\
    \text{FEVER} & \textbf{0.682} & 0.623 & 0.641 & \underline{0.670} \\
    \text{Climate-FEVER} & 0.187 & \textbf{0.211} & 0.200 & \underline{0.205} \\
    \text{SciFact} & 0.533 & \underline{0.554} & 0.523 & \textbf{0.588} \\
    \hline
    AVERAGE & \underline{0.370} & 0.359 & 0.360& \textbf{0.395} \\
\hline
\end{tabular}
\end{adjustbox}
    \caption{Zero-shot retrieval performances on \textbf{BEIR} benchmark (measured by NDCG@10).}
\label{tab:beir}

\end{table}